\documentclass[twocolumn,preprintnumbers,amsmath,amssymb]{revtex4-1}

\usepackage{array}
\usepackage[dvipdfmx]{graphicx}
\usepackage{tabularx}
\usepackage{longtable}
\usepackage{dcolumn} 
\usepackage{bm}
\usepackage{color}

\begin{document}

\title{A minimal electrostatic theory for the Seebeck coefficient in liquids}

\author{Wataru Kobayashi}
\email{kobayashi.wataru.gf@u.tsukuba.ac.jp}
\affiliation
{Division of Physics, Faculty of Pure and Applied Sciences, University of Tsukuba, 
Ibaraki 305-8571, Japan}

\begin{abstract}

The Seebeck coefficient in liquids often reaches the mV/K range, yet its microscopic origin remains 
unclear due to the complexity of electrolyte systems. Here we propose a minimal electrostatic 
theory focusing on solvation entropy. Using a core-shell extension of the Born equation with temperature-dependent 
dielectric response, we provide an alternative electrostatic interpretation of the mV/K-scale solvent-structuring 
contribution extracted for a cobalt redox electrolyte. 
The theory clarifies that large valence, small cationic radius, small dielectric constant, 
and large magnitude of $\frac{d\varepsilon}{dT}$ 
are key factors for enhanced liquid Seebeck response.

\end{abstract}

\maketitle
\section{Introduction}

The Seebeck effect is fundamental in condensed matter physics. 
In solids, the Seebeck coefficient ($S_{\rm e}$) is well interpreted by several methods such as 
Boltzmann equation \cite{singh}, linear response theory \cite{matsuura}, and 
Heikes formula \cite{chaikin,koshibae}. 
In liquids, $S_{\rm e}$ is also observed, and is found to be of the order of mV/K \cite{quickenden}. 
Compared with solids, the microscopic origin of $S_{\rm e}$ in liquids is still unclear.
This is because measurements of $S_{\rm e}$ in liquids require complex electrolyte systems including 
cathode/anode with cationic/anionic species. 

The liquid Seebeck effects in ionic and redox electrolytes are generally classified into thermodiffusive 
and thermogalvanic contributions \cite{jabeen, li, kang}. The former originates from ion redistribution under a temperature gradient, 
whereas the latter is governed by the entropy change of the electrode redox reaction. 
For thermogalvanic cells under open-circuit and near-equilibrium conditions, the Seebeck 
coefficient is related to the molar reaction entropy (in Jmol$^{-1}$K$^{-1}$) 
$\Delta S_{\rm rxn, mol}$ as $S_{\rm e}=\frac{\Delta S_{\rm rxn, mol}}{nF}$, 
where $n$ is a number of reaction electrons, and $F$ is Faraday constant \cite{jabeen}. 
Because the reaction entropy of simple redox couples contains a large solvation-entropy contribution, 
Born-type electrostatic models have long provided a useful thermodynamic reference 
for understanding the effects of ionic charge, effective radius, and solvent dielectric response. 
Weaver and coworkers used the Born dielectric medium model to quantitatively interpret 
solvation entropies in many transition-metal redox couples \cite{sahami,hupp}. 
However, they predicted substantially lower values than the experimental values. 
This discrepancy may be attributed to the fact that the classical Born model treats the solvent as a uniform bulk dielectric continuum and 
therefore does not explicitly distinguish the first solvation shell from the outer bulk solvent. 

Recent molecular-dynamics simulations have shown that the thermopower of liquid redox pairs 
is strongly affected by the local solvation-shell structure, including solvent dipole orientation 
and its fluctuation in the first solvation shell \cite{chen}. These results suggest that the first solvation 
region can have a local electrostatic and entropic response different from that of the bulk solvent. 
Motivated by this microscopic picture, the present work introduces a core-shell extension of 
the Born model, in which the local solvation shell and the outer bulk solvent are described by 
different effective dielectric responses.
The purpose of this model is not to replace atomistic simulations or full transport theories. 
Rather, it provides a minimal analytical bridge between the classical uniform-dielectric Born model 
and microscopic solvation-shell descriptions. In this sense, the present core-shell Born model 
should be regarded as a lowest-order analytical extension for incorporating first-solvation-shell 
effects into the electrostatic solvation-entropy contribution to the liquid Seebeck coefficient.

We apply this treatment to interpret $S_{\rm e}$ value of 
redox couples of cobalt complexes in $\gamma$-butyrolactone (GBL), although 
Cho {\it et al.} have already succeeded in analyzing the $S_{\rm e}$ value by using another method \cite{cho}. 
Combining viscosity measurements and quantum-chemical calculations, 
they quantitatively explained $S_{\rm e}$ in liquids as a sum of $S_{\rm e,ele}$, 
$S_{\rm e,vib}$, and $S_{\rm e,str}$, where $S_{\rm e,ele}$, 
$S_{\rm e,vib}$, and $S_{\rm e,str}$ represent electronic, vibrational, 
and solvent-structuring contributions of the Seebeck coefficient, respectively. 
In the following, $S_{\rm e,str}$ denotes the solvent-structuring contribution extracted in Ref. \cite{cho}, 
whereas $S_{\rm e,solv}$ denotes the electrostatic solvation-entropy contribution calculated 
using the present Born-type model. 
These two quantities are physically related but are not assumed to be strictly identical.

Here, we propose a core-shell extension of the Born dielectric model, in which the temperature-dependent dielectric 
responses of the first solvation shell and the bulk solvent are treated separately. 
The effective ionic radii and the reference value $S_{\rm e,str}=0.914$ mV/K used in the present analysis are 
adopted from Ref. \cite{cho}. The purpose of the present model is to examine whether the magnitude of 
this extracted solvent-structuring contribution can be represented by an effective electrostatic solvation response. 
The resulting agreement should therefore be regarded as an alternative electrostatic interpretation of $S_{\rm e,str}$, 
rather than as an independent reproduction of a directly measured solvation contribution.

\section{methods}

One may evaluate the solvent entropies using microscopic quantum-statistical calculation 
as a first-principle method. However, such a calculation would be unrealistic, 
because grand partition function of the electrolyte systems containing freely moving cationic and 
anionic species, solute ions/molecules, and solvent molecules is highly difficult to calculate 
for their complexity. 
To avoid this situation, we can simply use macroscopic electrostatics and thermodynamics 
to construct the theory of $S_{\rm e,solv}$ in liquids. 

\begin{figure}[t]
\begin{center}
\includegraphics[width=85mm,trim=20 400 20 90,clip]{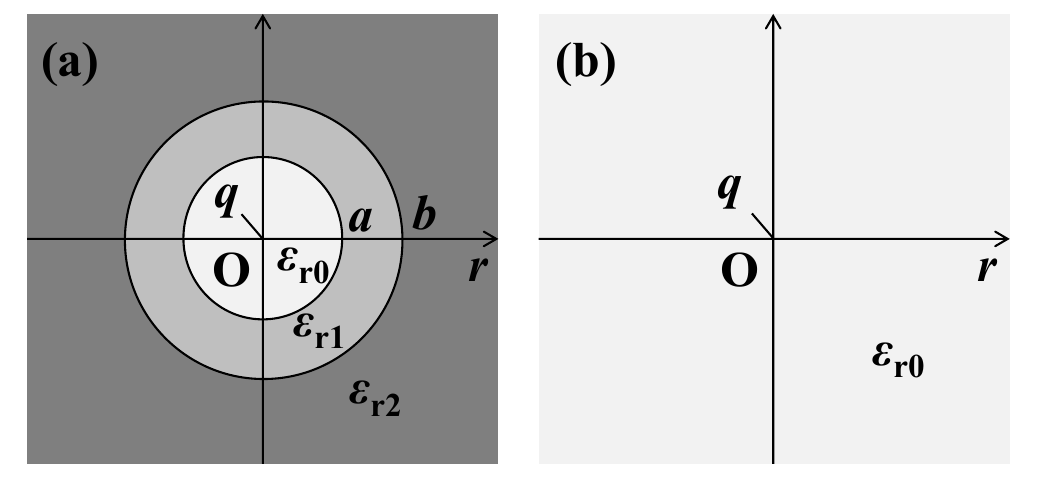}
\caption{(a) Schematic figure of core-shell structure in dielectric medium. Radius of the core 
and the shell are $a$ and $b$ m, respectively. Relative dielectric constant 
($\varepsilon_{{\rm r}i} (i=0,1,2)$) of the core, the shell, and the medium is 
$\varepsilon_{{\rm r}0}$, $\varepsilon_{{\rm r}1}$, and $\varepsilon_{{\rm r}2}$, respectively. 
The core represents a cation with a point charge $q$ at original point O. 
The shell and the medium represent a solvation structure and solvent, respectively. 
(b) Schematic figure of cation gas. A point charge $q$ at O is embedded in a medium with 
$\varepsilon_{{\rm r}0}$. This situation represents that a core is not in the solvent but is isolated. 
}
\label{f1}
\end{center}
\end{figure} 

\subsection{Assumption}
In our model, cations, anions, and solvent molecules compose electrolyte. 
One can measure $S_{\rm e}$ of the electrolyte using thermocell 
in which two platinum wires or plates as electrodes 
and bridged two electrolyte-filled beaker cells with different temperatures \cite{cho}. 
Ref. \cite{cho} does not explicitly report a correction or calibration for the thermoelectric contributions 
of the Pt electrodes and associated voltage leads. Since the absolute Seebeck coefficient of Pt is approximately $-5$ $\mu$V/K 
near room temperature\cite{kockert}, the Pt-wire contribution alone is expected to be on the order of several $\mu$V/K, 
corresponding to less than approximately 1\% of the adopted reference value of 0.914 mV/K. 
However, the exact magnitude and sign of the combined electrode and lead contribution cannot be 
determined without the detailed wiring configuration and calibration procedure. 
This possible instrumental contribution and the measurement uncertainty of Ref. \cite{cho} are 
inherited by the fitted apparent shell dielectric parameter. 
We assume that the cation is a sphere (core) with radius of $a$ m. The core has 
a valency of $q$ [$q=Ze$, $Z:$ oxidation number ($Z=+2$ or $+3$ in this work), $e:$ electron charge 
($e=1.602\times10^{-19}$ C)] at an original point O as shown in Fig. 1(a). 
This cation core is surrounded by a solvation shell with thickness of $\delta$ m 
($b \equiv a+\delta$). And the shell is further surrounded by solvent molecules as medium. 
In this situation, only dielectric constant represent physical characteristics of 
the electrolytes. 

The spherical core-shell geometry should be interpreted as a coarse-grained electrostatic approximation 
rather than as a literal molecular structure. The actual redox complex and the surrounding solvent molecules 
are not spherical, and the first solvation shell is anisotropic, heterogeneous, and fluctuating at the molecular level. 
In the present continuum model, these microscopic degrees of freedom are projected onto an orientationally 
averaged radial dielectric response. Thus, the spherical shell represents the lowest-order isotropic 
coarse-graining of the first solvation region.
Similarly, the shell dielectric constant should not be regarded as a uniquely determined microscopic 
dielectric constant of a geometrically well-defined shell. It is an apparent effective parameter representing 
the averaged local solvation response around the redox ion. This effective response may include reduced 
orientational polarization near the multivalent ion, specific solute-solvent interactions, solvent-mediated correlations, 
ion-pairing effects, and activity-coefficient corrections that are not explicitly resolved in the present continuum model 
(We further consider these effects in section III. Discussion.). 
A more microscopic treatment would require molecular dynamics simulations, local dielectric-response calculations, 
or an extension to an ellipsoidal or anisotropic dielectric shell.

The core, shell, and solvent exhibit $\varepsilon_{\rm 0}$, $\varepsilon_{\rm 1}$, and 
$\varepsilon_{\rm 2}$, respectively. By using Gauss' law, a magnitude of radial electric field ($E_{\rm r}$) 
is described as  
\begin{equation}
E_{\rm r}(r)=\frac{q}{4\pi \varepsilon_0 \varepsilon_{{\rm r}i} r^2}, 
\label{1}
\end{equation}
where $\varepsilon_{{\rm r}i}$ ($i=0,1,2$) is relative dielectric constant with 
$\varepsilon_i= \varepsilon_{{\rm r}i}\varepsilon_0$ [$\varepsilon_0:$ dielectric constant 
in vacuum ($\varepsilon_0=8.854\times 10^{-12}$ F/m)]. 
Then, the electrostatic potential $\phi(r)$ is defined as 
\begin{equation}
\phi (r) \equiv \int^{\infty}_r E_r(r') dr'. 
\label{2}
\end{equation}
Thus, at original point O, $\phi(0)$ is described as 
\begin{equation}
\begin{split}
\phi (0) &= \int^a_0 \frac{q}{4\pi \varepsilon_0 \varepsilon_{{\rm r}0} r^2} dr\\
&+\int^b_a \frac{q}{4\pi \varepsilon_0 \varepsilon_{{\rm r}1} r^2} dr+
\int^{\infty}_b \frac{q}{4\pi \varepsilon_0 \varepsilon_{{\rm r}2} r^2} dr. 
\label{3}
\end{split}
\end{equation}
The divergence at $r \rightarrow 0$ will cancel out in the reaction potential shown below. 
Reaction potential is defined as a difference in between solvation potential and reference potential 
(The reaction potential represents the change in electrostatic potential relative to the gas-phase reference; 
the corresponding electrostatic contribution to the solvation Gibbs free energy is introduced below.). 
The reference potential is thought to be 
\begin{equation}
\phi_{\rm ref} (0) \equiv \int^{\infty}_0 \frac{q}{4\pi \varepsilon_0 \varepsilon_{{\rm r}0} r^2} dr, 
\label{4}
\end{equation}  
as shown in Fig. 1(b). This $\phi_{\rm ref} (0)$ corresponds to gas phase, and 
thus $\varepsilon_{{\rm r}0}= 1$ in vacuum is assumed.  
Thus, the reaction potential 
$\phi_{\rm reac}(0)$ is determined by $\phi(0)-\phi_{\rm ref}(0)$ as 
\begin{equation}
\begin{split}
\phi_{\rm reac}(0) &\equiv \phi(0)-\phi_{\rm ref}(0)\\
&= \int^b_a \frac{q}{4\pi \varepsilon_0}\frac{dr}{r^2}\left(\frac{1}{\varepsilon_{{\rm r}1}}-
\frac{1}{\varepsilon_{{\rm r}0}}\right)\\
&+\int^{\infty}_b \frac{q}{4\pi \varepsilon_0}\frac{dr}{r^2}\left(\frac{1}{\varepsilon_{{\rm r}2}}-
\frac{1}{\varepsilon_{{\rm r}0}}\right)\\
&=\frac{q}{4\pi \varepsilon_0}\left[\frac{1}{a}\left(\frac{1}{\varepsilon_{{\rm r}1}}-
\frac{1}{\varepsilon_{{\rm r}0}}\right)+\frac{1}{b}\left(\frac{1}{\varepsilon_{{\rm r}2}}-
\frac{1}{\varepsilon_{{\rm r}1}}\right)\right].
\label{5}
\end{split}
\end{equation}  

Here, according to the electrostatics, electrostatic charging energy ($U_{\rm el}$) of the system shown in 
Fig. 1(a) is written as 
\begin{equation}
U_{\rm el} = \frac{1}{2}\int \rho(\bm{r})\phi(\bm{r})d^3\bm{r}, 
\label{6}
\end{equation}  
where $\rho(\bm{r})$ represents charge density. 
When $\rho(\bm{r})=q\delta(\bm{r})$, then Eq. \ref{6} reduces to $U_{\rm el}=\frac{1}{2}q\phi(0)$. 
The same expression is obtained from a reversible charging process. 
When a charge is quasistatically increased from zero to $q$, the required charging work is 
\begin{equation}
W^{\rm rev}_{\rm ch} = \int^q_0 \phi(q')dq'=\frac{1}{2}q\phi(q), 
\label{7}
\end{equation}  
using $\phi \propto \frac{q}{r}$. Under constant-temperature and constant-pressure conditions, 
this reversible charging work can be identified with the electrostatic contribution to the solvation Gibbs free-energy change 
$\Delta G^{\rm el}_{\rm solv}$. 
Therefore, relative to the reference medium, we define $\Delta G^{\rm el}_{\rm solv}$ as, 
\begin{equation}
\begin{split}
\Delta G^{\rm el}_{\rm solv} &\equiv \frac{1}{2}q(\phi(0)-\phi_{\rm ref}(0))=\frac{1}{2}q\phi_{\rm reac}(0)\\
&=\frac{q^2}{8\pi\varepsilon_0}\left[\frac{1}{a}\left(\frac{1}{\varepsilon_{{\rm r}1}}-
\frac{1}{\varepsilon_{{\rm r}0}}\right)+\frac{1}{b}\left(\frac{1}{\varepsilon_{{\rm r}2}}-
\frac{1}{\varepsilon_{{\rm r}1}}\right)\right].  
\label{8}
\end{split}
\end{equation}  
This is so called, the extended Born equation. When the thickness of the shell ($\delta$) is 
zero ($\delta=0 \rightarrow b=a$), this equation reduces to the Born equation as 
\begin{equation}
\Delta G^{\rm el}_{\rm solv} =\frac{q^2}{8\pi\varepsilon_0}\left[\frac{1}{a}\left(\frac{1}{\varepsilon_{{\rm r}2}}-
\frac{1}{\varepsilon_{{\rm r}0}}\right)\right].  
\label{9}
\end{equation}  

\subsection{Born equation}
Let us evaluate a value of solvation contribution $S_{\rm e,solv}$ of the Seebeck coefficient. 
First, we use a simple Born equation Eq. \ref{9} (a condition of $a=b$). 
We would like to note that cations and anions are independently moving in solvent. 
Namely, we treat non-interacting-free-cation-particle model with core-shell structure. 
We are assuming $\varepsilon_{{\rm r}0} =1$. Because the valence charges in the cation are strongly 
confined in the atomic core, and the charges cannot play as a dipole moment. 

Now, we introduce temperature dependence in relative dielectric constant ($\varepsilon_{{\rm r}i}(T)$). 
Then, Eq. \ref{9} becomes
\begin{equation}
\Delta G^{\rm el}_{\rm solv} =-\frac{q^2}{8\pi\varepsilon_0}\left[\frac{1}{a}\left(1-\frac{1}{\varepsilon_{{\rm r}2}(T)}\right)\right].  
\label{10}
\end{equation}  
The temperature derivative of $\Delta G^{\rm el}_{\rm solv}$ becomes entropy ($S_{\rm solv}$) as 
\begin{equation}
\begin{split}
S_{\rm solv}&=-\left(\frac{\partial \Delta G^{\rm el}_{\rm solv}}{\partial T}\right)_p\\
&=\frac{q^2}{8\pi\varepsilon_0a}\frac{1}{\varepsilon_{{\rm r}2}(T)^2}\frac{d\varepsilon_{{\rm r}2}(T)}{dT}.  
\label{11}
\end{split}
\end{equation} 
Since $\frac{d\varepsilon_{{\rm r}2}(T)}{dT}$is generally negative, $S_{\rm solv}<0$ is expected.  

\begin{table*}[t]
\centering
\caption{Summary of physical and structural parameters in a redox couple of 
cobalt complexes Co(bpy)$_3^{2+}$/Co(bpy)$_3^{3+}$ (bpy : bipyridine) in GBL using 
the Born model.}
\begin{tabular}{lll}
parameter & value & reference \\
\hline
\hline
core radius $a_{\rm ox}$ & $6.009 \times 10^{-10}$ m & Ref. \cite{cho} \\
core radius $a_{\rm red}$ & $6.169 \times 10^{-10}$ m & Ref. \cite{cho} \\
$\varepsilon_{{\rm r}2}$ of GBL & 42.82 at 293.15 K & Ref. \cite{fornefeld} \\
$\frac{d\varepsilon_{{\rm r}2}(T)}{dT}$ &  $-0.1455$ K$^{-1}$& Ref. \cite{fornefeld} \\
$\frac{1}{\varepsilon_{{\rm r}2}(T)^2}\frac{d\varepsilon_{{\rm r}2}(T)}{dT}$ &  $-7.94 
\times 10^{-5}$ K$^{-1}$& this work \\
oxidation number $Z_{\rm ox}$ & 3 & Ref. \cite{cho} \\
reduction number $Z_{\rm red}$ & 2 & Ref. \cite{cho} \\
theoretical $S_{\rm e,solv}^{\rm Born}$ & 0.486 mV/K & this work \\
solvent-structuring contribution $S_{\rm e,str}$ extracted from Ref. \cite{cho} & 0.914 mV/K & Ref. \cite{cho} \\
\hline
\end{tabular}
\end{table*}

Here, we introduce redox reaction as 
\begin{equation}
{\rm Ox}+ne^- \rightleftarrows {\rm Red}.  
\label{12}
\end{equation}  
Ox and Red represent oxidation (e.g. 3+) and reduction (e.g. 2+) states, respectively. 
$n$ is a number of reaction electrons. In this case, reaction entropy is expressed as 
(reaction $\rightarrow$ rxn)
\begin{equation}
\begin{split}
\Delta S_{\rm rxn}^{\rm Born}(T)&\equiv  S_{\rm solv,red}-S_{\rm solv,ox}\\
&=\frac{1}{8\pi\varepsilon_0}\frac{1}{\varepsilon_{{\rm r}2}(T)^2}\frac{d\varepsilon_{{\rm r}2}(T)}{dT}
\left(\frac{q_{\rm red}^2}{a_{\rm red}}-\frac{q_{\rm ox}^2}{a_{\rm ox}}\right).  
\label{13}
\end{split}
\end{equation} 
Since $q=Ze$, reaction entropy per mol ($N_{\rm A}:$ Avogadro number) becomes
\begin{equation}
\Delta S_{\rm rxn,mol}^{\rm Born}(T)=\frac{e^2N_{\rm A}}{8\pi\varepsilon_0}\frac{1}{\varepsilon_{{\rm r}2}(T)^2}
\frac{d\varepsilon_{{\rm r}2}(T)}{dT}\left(\frac{Z_{\rm red}^2}{a_{\rm red}}-\frac{Z_{\rm ox}^2}{a_{\rm ox}}\right).  
\label{14}
\end{equation} 
Seebeck coefficient $S_{\rm e,solv}^{\rm Born}$ is related to the reaction entropy as 
\begin{equation}
\begin{split}
S_{\rm e,solv}^{\rm Born}(T)&=\frac{\Delta S_{\rm rxn,mol}^{\rm Born}(T)}{nF}=
\frac{\Delta S_{\rm rxn,mol}^{\rm Born}(T)}{neN_{\rm A}}\\
&=\frac{e}{8\pi\varepsilon_0 n}\frac{1}{\varepsilon_{{\rm r}2}(T)^2}\frac{d\varepsilon_{{\rm r}2}(T)}{dT}
\left(\frac{Z_{\rm red}^2}{a_{\rm red}}-\frac{Z_{\rm ox}^2}{a_{\rm ox}}\right).  
\label{15}
\end{split}
\end{equation} 

Now, we can quantitatively calculate a value of $S_{\rm e,solv}$. 
Note that in the simple Born estimate, all parameters in Eq. \ref{15} can be determined from experimental 
or literature values without an adjustable shell parameter.
Table I summarizes physical and structural parameters of a redox couple of 
cobalt complexes Co(bpy)$_3^{2+}$/Co(bpy)$_3^{3+}$ (bpy : bipyridine) 
in $\gamma$-butyrolactone (GBL) \cite{cho}. 
 
By using $n=1$, $\frac{1}{\varepsilon_{{\rm r}2}(T)^2}\frac{d\varepsilon_{{\rm r}2}(T)}{dT}=
-7.94\times 10^{-5}$K$^{-1}$\cite{fornefeld}, 
$Z_{\rm red}=2$, $Z_{\rm ox}=3$, $a_{\rm red}=6.169 \times 10^{-10}$ m \cite{cho}, 
$a_{\rm ox}=6.009 \times 10^{-10}$ m \cite{cho}, 
$S_{\rm e,solv}^{\rm Born}(T)$ was evaluated to be 0.486 mV/K for a Co(bpy)$_3^{2+/3+}$ couple. 
Despite the simplicity of the model, the parameter-free Born estimate has the same mV/K order 
of magnitude as the solvent-structuring contribution extracted in Ref. \cite{cho}. 
This suggests that electrostatic solvation entropy provides a major mV/K-scale contribution 
to the thermogalvanic response of the present liquid redox system. 
Next, we consider a possible origin of the difference. 

\begin{table*}[t]
\centering
\caption{Summary of physical and structural parameters in a redox couple of 
cobalt complexes Co(bpy)$_3^{2+}$/Co(bpy)$_3^{3+}$ in GBL using 
the extended Born model.}
\begin{tabular}{lll}
parameter & value & reference \\
\hline
\hline
shell thickness $\delta$ & $3 \times 10^{-10}$ m & this work \\
shell radius $b_{\rm ox}$ & $9.009 \times 10^{-10}$ m & this work \\
shell radius $b_{\rm red}$ & $9.169 \times 10^{-10}$ m & this work \\
fitted value of $\varepsilon_{{\rm r}1}$ & 22.5 & this work \\
assumed $\frac{d\varepsilon_{{\rm r}1}(T)}{dT}$ &  $-0.1455$ K$^{-1}$& this work \\
calculated $\frac{1}{\varepsilon_{{\rm r}1}(T)^2}\frac{d\varepsilon_{{\rm r}1}(T)}{dT}$ &  $-2.87 \times 10^{-4}$ K$^{-1}$& this work \\
fitted $S_{\rm e,solv}^{\rm exBorn}$ & 0.914 mV/K & this work \\
solvent-structuring contribution $S_{\rm e,str}$  extracted from Ref. \cite{cho} & 0.914 mV/K & Ref. \cite{cho}\\
\hline
\end{tabular}
\end{table*}

\subsection{Extended Born equation}
We use the extended Born equation to evaluate the value of $S_{\rm e, solv}$. 
Similarly, the expression of $S_{\rm e,solv}^{\rm exBorn}(T)$ is obtained from Eq. \ref{8} as 
\begin{equation}
\begin{split}
S_{\rm e,solv}^{\rm exBorn}(T)&=\frac{\Delta S_{\rm rxn,mol}^{\rm exBorn}(T)}{nF}=
\frac{\Delta S_{\rm rxn,mol}^{\rm exBorn}(T)}{neN_{\rm A}}\\
&=\frac{e}{8\pi\varepsilon_0 n}\frac{1}{\varepsilon_{{\rm r}1}(T)^2}\frac{d\varepsilon_{{\rm r}1} (T)}{dT}\\
&\times\left[Z_{\rm red}^2\left(\frac{1}{a_{\rm red}}-\frac{1}{b_{\rm red}}\right)-Z_{\rm ox}^2\left(\frac{1}{a_{\rm ox}}-\frac{1}{b_{\rm ox}}\right)\right]\\
&+\frac{e}{8\pi\varepsilon_0 n}\frac{1}{\varepsilon_{{\rm r}2}(T)^2}\frac{d\varepsilon_{{\rm r}2} (T)}{dT}
\left(\frac{Z_{\rm red}^2}{b_{\rm red}}-\frac{Z_{\rm ox}^2}{b_{\rm ox}}\right),  
\label{16}
\end{split}
\end{equation} 
where $\varepsilon_{{\rm r}1}$ represents an effective dielectric constant of the first solvation shell, 
which is commonly introduced in continuum solvation models.
The shell thickness was assumed to be $3\times10^{-10}$m, 
which corresponds to the molecular radius of $\gamma$-butyrolactone and 
represents the first solvation shell. 
By using $n=1$, $\frac{1}{\varepsilon_{{\rm r}2}(T)^2}\frac{d\varepsilon_{{\rm r}2} (T)}{dT}=
-7.94 \times 10^{-5}$K$^{-1}$ \cite{fornefeld}, 
$Z_{\rm red}=2$, $Z_{\rm ox}=3$, $b_{\rm red}=9.169 \times 10^{-10}$ m, and 
$b_{\rm ox}=9.009 \times 10^{-10}$ m as listed in Table II, 
the bulk contribution, corresponding to the second term on the right-hand side of Eq. \ref{16}, was evaluated to be 0.322 mV/K. 
The remaining 0.592 mV/K of the solvent-structuring contribution extracted in Ref. \cite{cho} was therefore assigned to the 
effective first-solvation-shell contribution. Because Eq. \ref{16} contains the shell dielectric response through the combination 
$\frac{1}{\varepsilon_{{\rm r}1}^2}\frac{d\varepsilon_{{\rm r}1}(T)}{dT}$, 
the values of $\varepsilon_{{\rm r}1}$ and $\frac{d\varepsilon_{{\rm r}1}(T)}{dT}$ cannot be determined independently 
from the reference value alone. To reduce the number of undetermined parameters, 
we assume that the absolute temperature derivative of the apparent shell dielectric response is equal to that of the bulk solvent, 
$\frac{d\varepsilon_{{\rm r}1}(T)}{dT}=\frac{d\varepsilon_{{\rm r}2}(T)}{dT}=-0.1455$ K$^{-1}$. 
Under this assumption, the extended Born contribution $S_{\rm e,solv}^{\rm exBorn}(T)$ becomes 0.914 mV/K for $\varepsilon_{{\rm r}1}=22.5$ 
using $a_{\rm red}=6.169 \times 10^{-10}$ m \cite{cho}, and 
$a_{\rm ox}=6.009 \times 10^{-10}$ m \cite{cho}. 
The fitted value $\varepsilon_{{\rm r}1}=22.5$ is therefore conditional on the assumed temperature derivative and 
should be regarded as an apparent effective shell dielectric constant, 
rather than as an independently determined microscopic dielectric constant. 
The fitted value of $\varepsilon_{{\rm r}1}$ ($=22.5$) is almost half of the value of 
$\varepsilon_{{\rm r}2}$ ($=42.82$). 
The smaller fitted value of $\varepsilon_{{\rm r}1}$ compared with the bulk value $\varepsilon_{{\rm r}2}=42.82$ may 
be interpreted as an apparent reduction of the local dielectric response in the first solvation region. 
Such a reduction is consistent with the physical picture that solvent dipoles near a multivalent ion are 
partially oriented or constrained and therefore do not respond as freely as in the bulk solvent. 
We note that the fitted value of $\varepsilon_{{\rm r}1}$ slightly depends on the value of 
$\delta$ as listed in Table III. 

\begin{table}[t]
\centering
\caption{Sensitivity of the fitted apparent shell dielectric constant $\varepsilon_{{\rm r}1}$ to the assumed shell 
thickness $\delta$. For each $\delta$, $\varepsilon_{{\rm r}1}$ was fitted so that the extended Born contribution equals 
the solvent-structuring contribution extracted in Ref. \cite{cho}, $S_{\rm e,str}=$0.914 mV/K. 
The fitted values remain below the bulk dielectric constant of GBL, $\varepsilon_{{\rm r}2}=42.82$.}
\begin{tabular}{ll}
shell thickness $\delta$(m) & fitted value of $\varepsilon_{{\rm r}1}$ \\
\hline
\hline
$2 \times 10^{-10}$ & 20.2 \\
$3 \times 10^{-10}$ & 22.5 \\
$4 \times 10^{-10}$ & 24.0 \\
$5 \times 10^{-10}$ & 25.0 \\
\hline
\end{tabular}
\end{table}

In the extended Born analysis, the fitted value $\varepsilon_{{\rm r}1}\sim22.5$ should be interpreted with caution. 
This value is obtained by assigning the remaining 0.592 mV/K, required to represent the solvent-structuring contribution 
extracted in Ref. \cite{cho} after subtracting the bulk contribution of 0.322 mV/K, to the effective shell response. 
Therefore, it is not a parameter-free determination of the microscopic dielectric 
constant of the first solvation shell. Rather, it is an apparent effective shell dielectric constant within the present 
level of coarse-graining.
If additional corrections, such as thermodiffusion-induced concentration gradients, activity-coefficient effects, 
ion pairing, or other non-Born contributions, are treated explicitly, the residual contribution assigned to the 
shell term would change. Consequently, the fitted $\varepsilon_{{\rm r}1}$ would also be modified. 
The present result should therefore be understood as showing that the solvent-structuring contribution extracted 
in Ref. \cite{cho} beyond the bulk contribution of the extended Born model can be represented by a physically plausible reduction 
of the local dielectric response relative to the bulk solvent.
Direct microscopic determination of the reduced local dielectric response requires future molecular-dynamics 
simulations or local dielectric-response calculations and is beyond the scope of the present work.

\section{discussion}
\subsection{Relation to previous theoretical, computational and experimental studies}

The present model can be positioned relative to previous theoretical, computational, and experimental 
studies of liquid thermogalvanic systems. 
As stated in Introduction, 
Weaver and coworkers used Born-type electrostatic models to analyze solvation entropies of transition-metal redox couples, 
showing that ionic charge, effective radius, and solvent dielectric response are important factors in reaction entropy. 

Recent thermocell studies have further clarified the limitation of the classical Born model based on 
a uniform dielectric continuum. For example, Inoue {\it et al.} showed that molecular-level solvent-ion association 
can enhance the Seebeck coefficient beyond the estimate of the classical Born model \cite{inoue}. 
This result indicates that the uniform-dielectric Born model should be regarded as an electrostatic reference 
limit and that first-solvation-shell structures can play an important role in thermogalvanic reaction entropy. 
Explicit molecular association effects are discussed separately in Sec. III C.

Cho {\it et al.} analyzed cobalt-complex thermogalvanic electrolytes in GBL by combining thermoelectrochemical measurements, 
viscosity analysis, and quantum-chemical calculations, and decomposed the Seebeck coefficient into electronic, vibrational, 
and solvent-structuring contributions. The present theory follows this thermodynamic line, but focuses specifically 
on the electrostatic solvation-entropy contribution and expresses it in a compact analytical form.

The solvent dependence of Fe$^{3+}$/Fe$^{2+}$ thermopower also supports the electrostatic interpretation. 
Chen {\it et al.} reported that the experimentally determined thermogalvanic temperature coefficient of the Fe$^{3+}$/Fe$^{2+}$ 
redox pair increases from $1.2\pm0.1$ mV/K in water to $3.9\pm0.3$ mV/K in acetone \cite{chen}. 
This corresponds to an enhancement by a factor of approximately four in the measured thermopower. 
On the other hand, using only the solvent dielectric factor appearing in the Born-type expression, 
$(1/\varepsilon_r^2)(d\varepsilon_r/dT)$, the contribution in acetone \cite{onimisi} is estimated to be about 
five times larger than that in water \cite{malmberg}. This semi-quantitative agreement indicates that the dielectric 
response of the solvent captures the dominant scale of the solvent dependence of the thermogalvanic Seebeck coefficient.

At the same time, the mixed water/acetone results of Chen {\it et al.} show that the thermopower 
does not vary linearly with the bulk solvent composition. Instead, it increases rapidly only at 
high acetone fractions, where acetone molecules enter the first solvation shell of Fe$^{2+}$. 
This behavior cannot be described by a simple uniform-bulk dielectric model alone. 
It suggests that the local solvation-shell environment, rather than only the macroscopic solvent composition, 
controls the detailed solvent dependence of the thermopower. The present core-shell Born model is 
therefore motivated as a minimal continuum extension that separates the local shell response 
from the outer bulk dielectric response.

As another example of the electrostatic trend in thermogalvanic redox systems, Buckingham {\it et al.} 
systematically investigated iron-based anionic redox couples formed by complexation of Fe$^{2+/3+}$ with 
polycarboxylate and polyaminocarboxylate ligands \cite{buckingham}. They showed that 
the parent Fe$^{2+}$/Fe$^{3+}$ chloride electrolyte prepared from FeCl$_2$ and FeCl$_3$ 
exhibits a positive Seebeck coefficient of $+1.1$ mV/K, whereas ligand complexation 
can invert the sign and produce large negative Seebeck coefficients. In particular, the optimized 
([Fe(NTA)$_2$]$^{3-/4-}$) redox couple exhibits $S_{\rm e}=-1.35$ mV/K, 
comparable to the widely used ([Fe(CN)$_6$]$^{3-/4-}$) redox couple.

Importantly, they analyzed the reaction entropy calculated from the Seebeck coefficient and discussed 
an approximately proportional relationship between $\Delta S_{\rm rc}$ and $(Z_{\rm ox}^2-Z_{\rm red}^2)/r$. 
This empirical trend is consistent with the leading charge-density dependence expected from a 
Born-type electrostatic solvation model. Although their systems involve additional complications 
such as ligand coordination, pH-dependent speciation, vibrational entropy, and kinetic effects, 
the observed correlation suggests that ionic charge, effective size, and electrostatic solvation remain 
important organizing parameters across different thermogalvanic redox electrolytes. 

Therefore, the present core-shell Born model should not be regarded as a complete quantitative theory for 
all liquid redox systems. Rather, it provides a minimal analytical reference framework for the electrostatic 
part of the solvation-entropy contribution. It is complementary to microscopic molecular-dynamics simulations, 
semiempirical solution-chemistry analyses, and transport theories of thermodiffusion and concentration polarization.

\subsection{Corrections beyond the extended Born model}

The present extended Born model should be regarded as an electrostatic reference model for the 
solvation-entropy contribution to the liquid Seebeck coefficient. In real thermogalvanic cells, 
several non-Born effects may also contribute to the experimentally observed thermovoltage. 
Here we briefly estimate their possible magnitude and clarify the scope of the present model. 

According to the Nernst equation \cite{lefrou}, the formal potential $E$ of the redox reaction can be expressed as 
\begin{equation}
E=E^0+\frac{RT}{nF}
\ln\frac{\alpha_{\rm ox}(T)}{\alpha_{\rm red}(T)},
\label{17}
\end{equation}
where $E^0$ is the standard potential, $\alpha_i(T)$ is the activity of species ($i=$ red, ox), and $R$ is the ideal gas constant. 
Under isothermal condition, the potentials at both cathode and anode are in equilibrium causing no potential difference $\Delta E=0$. 
If temperature difference $\Delta T$ is applied in between these electrodes, finite voltage difference is induced as 
$\Delta E=S_{\rm e}\Delta T$. The temperature derivative of the standard potential $E^0(T)$ gives the equilibrium 
reaction-entropy contribution discussed above. Here, we define the additional activity-dependent contribution as 
$S_{\rm e, act}\equiv S_{\rm e}-\frac{dE^0}{dT}$. 
Now, considering temperature dependence of the activity, the expression of $S_{\rm e, act}$ is derived from the 
Nernst equation as, 
\begin{equation}
\begin{split}
S_{\rm e, act} &=\frac{dE}{dT}-\frac{dE^0}{dT}\\
&=\frac{R}{nF}\ln\frac{c_{\rm ox}(T)}{c_{\rm red}(T)}+\frac{RT}{nF}\frac{d}{dT}\ln\frac{c_{\rm ox}(T)}{c_{\rm red}(T)}\\
&+\frac{R}{nF}\ln\frac{\gamma_{\rm ox}(T)}{\gamma_{\rm red}(T)}+\frac{RT}{nF}\frac{d}{dT}\ln\frac{\gamma_{\rm ox}(T)}{\gamma_{\rm red}(T)}, 
\label{18}
\end{split}
\end{equation}
where $c_i(T)$ and $\gamma_i(T)$ are the concentration and activity coefficient of the species $i$ ($i=$ red and ox), respectively. 
Note that $\alpha_i =\gamma_i c_i$. The first term of Eq. \ref{18} represents the contribution from a fixed deviation 
of the concentration ratio from unity. 
Here, $c_{\rm ox}$ and $c_{\rm red}$ denote the local concentrations of the oxidized and reduced redox species, respectively, 
not the counter-anion concentration. In a symmetric thermogalvanic electrolyte, these concentrations are usually prepared to be equal, 
$c_{\rm ox}/c_{\rm red}=1$, and the concentration-ratio terms vanish in the absence of concentration polarization or thermodiffusion. 
The difference in the ionic valence of ox and red does not by itself require different values of $c_{\rm ox}$ and $c_{\rm red}$, 
because charge neutrality is maintained by the counterions. This term $(R/F)\ln(c_{\rm ox}/c_{\rm red})$ only 
gives 0.86 $\mu$V/K for a 1\% deviation. 

The second term of Eq. \ref{18} gives a contribution of thermodiffusion (Soret effect). 
Soret coefficient $S_{{\rm T},i}$ is defined as $S_{{\rm T},i}\equiv -\frac{d}{dT}\ln c_i(T)$. 
Then the second term is rewritten as 
\begin{equation}
S_{\rm e, Soret} = \frac{RT}{nF}\left(S_{\rm T, red}-S_{\rm T, ox}\right).
\label{19}
\end{equation}
For a typical difference in Soret coefficients of the order of ($10^{-3}~{\rm K}^{-1}$) \cite{lecce}, 
this contribution is of the order of several tens of ($\mu{\rm V/K}$). 

The third and fourth terms of Eq. \ref{18} represent activity-coefficient corrections associated with ion-ion interactions. 
The third term arises from a finite activity-coefficient ratio at a given temperature, 
whereas the fourth term arises from the temperature dependence of this ratio. 
An extended Debye-Huckel-type estimate \cite{lefrou} gives 
\begin{equation}
\ln\gamma_i = -\frac{A z_i^2 \sqrt{I}}{1+B a_i\sqrt{I}},
\label{20}
\end{equation}
where $I$ is the ionic strength, and $a_i$ is an effective ion-size parameter of species $i$ ($i=$ red and ox). 
$A$ ($B$) is a complex parameter which has the unit of L$^{1/2}$mol$^{-1/2}$ (L$^{1/2}$mol$^{-1/2}$nm$^{-1}$) \cite{lefrou}. 
For pure water at 25$^{\circ}$C, the value of $A$ is 0.509 L$^{1/2}$mol$^{-1/2}$. And the product of $aB$ is 
evaluated to be about 1 \cite{lefrou}. 
When concentration $c$ is 0.01 M, $I$ is of the order of 0.01 M. Thus, 
one obtains the order-of-magnitude estimate
\begin{equation}
|S^{(0)}_{{\rm e},\gamma}|\sim\frac{R}{nF}\frac{A|z_{\rm ox}^2-z_{\rm red}^2|\sqrt{I}}{1+B a_{\rm ave}\sqrt{I}}, 
\label{21}
\end{equation}
where $a_{\rm ave}$ means an average of $a_{\rm red}$ and $a_{\rm ox}$. 
For a (3+/2+) redox couple, ($|z_{\rm ox}^2-z_{\rm red}^2|=5$). 
Therefore, depending on ionic strength and effective ion size, this activity-coefficient correction can reach several tens $\mu{\rm V/K}$. 

For the fourth term of Eq. \ref{18}, within an extended Debye-Huckel description, this derivative generally includes the explicit 
temperature dependences of the Debye-Huckel parameters $A$ and $B$, as well as the implicit temperature dependence 
through the ionic strength $I$ and possibly the effective ion-size parameters $a$. 
Since the required temperature-dependent activity-coefficient data are unavailable for the present 
cobalt redox electrolyte in GBL, this term cannot be quantitatively evaluated here and remains an unquantified correction.

However, for an order-of-magnitude estimate, we retain only the explicit temperature dependence of the Debye-Huckel coefficient $A(T)$, 
while the ionic strength $I$, the coefficient $B$, and the effective ion-size parameters are held constant. 
Under this approximation, 
\begin{equation}
|S^{(T)}_{{\rm e},\gamma}|=\frac{RT}{nF}\bigg|\frac{d}{dT}\ln\frac{\gamma_{\rm ox}(T)}{\gamma_{\rm red}(T)}\bigg|
=T\bigg|\frac{d\ln A(T)}{dT}\bigg||S^{(0)}_{{\rm e},\gamma}|.
\label{22}
\end{equation}
For the concentration-based Debye-Huckel coefficient, $A(T)\propto [\varepsilon_{\rm r}(T)T]^{-3/2}$ \cite{lefrou}. 
Using the dielectric data of GBL gives $\frac{d\ln A(T)}{dT}\simeq -2.0 \times 10^{-5}$ K$^{-1}$ near room temperature. 
The resulting temperature-dependent activity-coefficient contribution is therefore approximately 0.6\% of 
the static activity-coefficient correction (the third term) and is negligibly small on the present mV/K scale.

Electrode kinetics provides another possible non-ideal correction. 
Under ideal open-circuit equilibrium conditions, the net Faradaic current is zero and 
the Butler-Volmer overpotential vanishes. Therefore, electrode kinetics does not directly 
contribute to the equilibrium Seebeck coefficient. In real measurements, however, 
a small residual current may arise from finite input resistance, internal leakage paths, 
self-discharge, side reactions, or non-steady relaxation processes. 
If this residual current is denoted by $j_{\rm leak}$, the corresponding kinetic overpotential 
is estimated from the linearized Butler-Volmer equation as
\begin{equation}
\eta \sim \frac{RT}{nF}\frac{j_{\rm leak}}{j_0},
\label{23}
\end{equation}
where $j_0$ is the exchange current density. The corresponding correction is then of the order
\begin{equation}
|S_{\rm e, kin}|\sim \frac{RT}{nF\Delta T}\frac{|j_{\rm leak}|}{j_0}.
\label{24}
\end{equation}
This term vanishes in the ideal open-circuit limit and is expected to be small under near-equilibrium conditions.

Within the independent-ion Born approximation, the direct solvation contribution of a non-redox-active 
anion cancels from the reaction entropy. 
Because the anion with core-shell structure does not change their charge 
through the redox reaction. When cation-anion interaction is considered, 
the anion effect might give a small correction. 
As a simple estimate, the pair interaction energy of species $i$ ($i={\rm ox,red}$) 
with an anion of charge ($z_{\rm A}e$) may be written as

\begin{equation}
G_{{\rm pair},i}\simeq \frac{N_{\rm A} e^2}{4\pi\varepsilon_0\varepsilon_r}\frac{z_i z_{\rm A}}{r_{i{\rm A}}},
\label{25}
\end{equation}
where $r_{i{\rm A}}$ is the effective cation-anion separation. The corresponding change in the pair interaction free energy 
upon the redox reaction is
\begin{equation}
\Delta G_{\rm pair} \equiv G_{{\rm pair},{\rm red}}-G_{{\rm pair},{\rm ox}}=
\frac{(z_{\rm ox}-z_{\rm red})N_{\rm A} e^2}{4\pi\varepsilon_0\varepsilon_r r_{i{\rm A}}}, 
\label{26}
\end{equation}
where $z_{\rm A}$ is $-1$. 
The corresponding entropy contribution is
$\Delta S_{\rm pair}=-\frac{\partial \Delta G_{\rm pair}}{\partial T}$, and the corresponding Seebeck coefficient $S_{\rm e, pair}$ is 
obtained as $S_{\rm e, pair}=\frac{\Delta S_{\rm pair}}{nF}$. If the temperature dependence of the pair distance is neglected, 
$S_{\rm e, pair}$ is expressed as 
\begin{equation}
|S_{\rm e, pair}|= \frac{|z_{\rm ox}-z_{\rm red}|e}{4\pi\varepsilon_0 n r_{i{\rm A}} \varepsilon_r^2}\bigg|\frac{d\varepsilon_r}{dT}\bigg|.
\label{27}
\end{equation}
This term is an indirect anion contribution: it does not originate from a change in the anion solvation itself, but from the 
change in the cation-anion Coulomb interaction caused by the redox change of the cation charge. 
Since it is proportional to the difference between $z_{\rm red}/r_{{\rm red}A}$ and $z_{\rm ox}/r_{{\rm ox}A}$, 
its magnitude depends on the effective ion-pair distance and the degree of ion pairing. 
In dilute electrolytes or in weakly associated ion pairs, this contribution is expected to be 
a correction to the main electrostatic solvation-entropy term, but it may become non-negligible 
when the redox cation and counter anion form relatively close ion pairs. 
When $n=1$, $\frac{1}{\varepsilon_r^2}\frac{d\varepsilon_r}{dT}=-7.94\times 10^{-5}$ K$^{-1}$, and $r_{i{\rm A}}=1$ nm, 
$S_{\rm e, pair}$ is evaluated to be 114 $\mu$V/K. Under this evaluation, we assume complete ion pairing, however, 
for partial ion pairing, this value should be multiplied by the ion-pairing fraction resulting in several tens $\mu$V/K. 

Specific solute-solvent interactions and solvent-mediated correlations are also not explicitly 
resolved in the present continuum model. These effects correspond to molecular-level interactions 
between the redox ion and nearby solvent molecules, as well as indirect correlations between ions 
through solvent orientation, density fluctuation, and structural rearrangement. 
In the present coarse-grained description, their averaged influence may be partially included in 
the apparent effective shell dielectric constant. Thus, the fitted $\varepsilon_{r1}$ should not 
be regarded as a pure microscopic dielectric constant of the first solvation shell, but as an 
effective local dielectric response that may include such higher-order local solvation effects.

These estimates indicate that non-Born corrections are not necessarily negligible for quantitative prediction. 
However, under dilute, near-equilibrium conditions, they are typically expected to be of the order of 
several tens to $10^2~\mu{\rm V/K}$, which is smaller than the mV/K-scale electrostatic 
solvation-entropy contribution considered here. Therefore, the fitted value $\varepsilon_{r1}\sim22.5$ 
in the present extended Born analysis should be interpreted as an apparent value obtained when 
the remaining contribution beyond the bulk term of the extended Born model is assigned to the effective shell response. 
If non-Born corrections are explicitly separated, the residual contribution attributed to the shell 
term, and hence the fitted $\varepsilon_{r1}$, would be modified.

\subsection{Molecular association effects beyond the present model}

The present core-shell Born model treats the first solvation shell as an effective dielectric region and does not 
explicitly describe specific chemical association between redox species and solvent molecules. 
However, recent thermocell studies have shown that such molecular association can also produce large reaction-entropy contributions. 
For example, Inoue {\it et al.} reported that the Seebeck coefficient of the p-chloranil radical/dianion redox pair in acetonitrile is enhanced 
from $-1.3$ to $-2.6~{\rm mV/K}$ by the addition of ethanol \cite{inoue}. This enhancement was attributed to the formation and dissociation 
of local ethanol-ion pairs around the dianion, namely an order-disorder transition of the local solvation structure.

This mechanism should be distinguished from the concentration-gradient Nernst correction discussed above. 
In the association mechanism, the large thermovoltage originates from the temperature dependence of a specific 
reversible solvent-ion association equilibrium, and therefore from the association-entropy contribution to the 
redox reaction entropy. Such an effect is not included explicitly in the present simple core-shell Born model. 
Nevertheless, it illustrates that first-solvation-shell structures can strongly affect the thermogalvanic reaction entropy. 
In this sense, molecular association models and the present core-shell Born model describe different levels of 
first-solvation-shell physics: the former treats specific chemical binding explicitly, whereas the latter provides 
a minimal electrostatic coarse-graining of the local solvation response.

\section{conclusion}

In conclusion, we have theoretically examined Seebeck coefficient in liquids. 
The present minimal theory suggests that electrostatic solvation entropy can provide a major 
contribution to the mV/K-scale Seebeck coefficient in liquid redox electrolytes. 
A large Seebeck coefficient in liquids is realized by a combination of large ionic valence, 
small cationic radius, small dielectric constant, and large magnitude of temperature derivative of the dielectric constant. 
This minimal electrostatic theory provides a simple physical guideline for understanding 
thermoelectric effects in liquid electrolytes. 

\section{acknowledgment}
We would like to thank H. Kobayashi and S. Kobayashi for support.

\end{document}